\documentclass[12pt]{iopart}

\usepackage{iopams}  
\usepackage{epsfig}
\usepackage{graphicx}

\begin{document}

\title[Coulomb repulsion and correlation strength in LaFeAsO]{Coulomb repulsion and correlation strength in LaFeAsO from Density Functional and Dynamical Mean-Field Theories}

\author {V.~I.~Anisimov$^1$,  Dm.~M.~Korotin$^1$, M.~A.~Korotin$^1$, A.~V.~Kozhevnikov$^{1,2}$, J.~Kune\v{s}$^3$, A.~O.~Shorikov$^1$, S.L.~Skornyakov$^1$ and S.~V.~Streltsov$^1$}

\address{$^1$Institute of Metal Physics, Russian Academy of Sciences,
620041 Yekaterinburg GSP-170, Russia\\
$^2$Joint Institute for Computational Sciences, Oak Ridge
National Laboratory P.O. Box 2008 Oak Ridge, TN 37831-6173, USA\\
$^3$Theoretical Physics III, Center for Electronic Correlations
and Magnetism, Institute of Physics, University of Augsburg, Augsburg
86135, Germany}

\begin {abstract}

LDA+DMFT (Local Density Approximation combined with Dynamical
Mean-Field Theory) computation scheme  has been
used to calculate spectral properties of LaFeAsO -- the parent compound
for new high-T$_c$ iron oxypnictides. Coulomb repulsion $U$ and Hund's
exchange $J$  parameters for iron $3d$ electrons  were calculated using \textit {first
principles} constrained density functional theory scheme in Wannier
functions formalism. Resulting values strongly depend on the number of states taken into account in
calculations: when full set of O-$2p$, As-$4p$, and Fe-$3d$ orbitals with corresponding
bands are included, computation results in $U=$3$\div$4~eV and J=0.8~eV. In contrast to that when
the basis set is restricted to Fe-$3d$ orbitals and bands only, computation
gives  much smaller parameter values $F^0$=0.8~eV, $J$=0.5~eV.  However, DMFT calculations
with both parameter sets and corresponding to them choice of basis
functions result in weakly correlated electronic structure that is in
agreement with experimental X-ray and photoemission spectra.

\end {abstract}

\pacs {74.25.Jb, 71.45.Gm}

\maketitle

Recent discovery of high-$T_c$ superconductivity in iron oxypnictides
LaO$_{1-x}$F$_x$FeAs~\cite {Kamihara-08} has stimulated
an intense experimental and theoretical activity. In striking similarity with high-$T_c$ cuprates,
undoped material LaFeAsO  is not superconducting with antiferromagnetic
commensurate spin density wave developing below 150~K~\cite {neutrons}. 
Only when electrons (or holes) are added to the system via doping,
antiferromagnetism is suppressed and superconductivity appears. As it is
generally accepted that Coulomb correlations between copper $3d$ electrons
are responsible for cuprates anomalous properties, it is tempting to
suggest that the same is true for iron $3d$ electrons in LaFeAsO.

\begin {figure}
\includegraphics [width=0.425\textwidth]{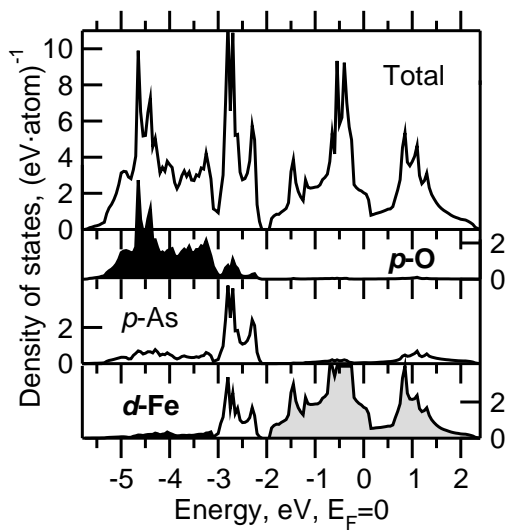}
\caption {Total and partial densities of states for LaFeAsO obtained in DFT
calculation in frame of LMTO method.}
\label {fig1} 
\end {figure}

The ratio of Coulomb interaction parameter $U$ and band width $W$
determines correlation strength. If $U/W<1$  then
the system is weakly correlated and results of the Density Functional
Theory (DFT) calculations are reliable enough to explain its electronic and
magnetic properties. However, if $U$ value is comparable
with $W$ or even larger then the system is in intermediate or
strongly correlated regime and Coulomb interactions must be explicitly
treated in electronic structure calculations. The partially filled bands formed by Fe-$3d$ states in LaFeAsO
have the width $\approx$4~eV (see shaded area in the lower panel of
Fig.~\ref {fig1}), so the estimation for Coulomb interaction parameter $U$ should be compared
with this value.

In practical calculations, $U$ is often considered as a free parameter to
achieve the best agreement of calculated and measured properties of
investigated system. Alternatively, $U$ value could be estimated from the
experimental spectra. However, the most attractive approach is to
determine Coulomb interaction parameter $U$ value in \textit {first principles} non-empirical way. There are two such methods: constrained DFT 
scheme~\cite {U-calc,anigun}, where  $d$-orbital
occupancies in DFT calculations are fixed to the certain values and $U$ is numerically
determined as a derivative of $d$-orbital energy over its occupancy, and 
Random Phase Approximation (RPA) method~\cite {RPA}, where screened Coulomb interaction
between $d$-electrons is calculated via perturbation theory. Recently, the RPA calculations for Coulomb
interaction parameter $U$ in LaFeAsO were reported~\cite {Arita}, where $U$
value was estimated as 1.8$\div$2.7 eV.  In 
Ref.~\cite {Haule08} it was proposed to use in LaFeAsO $U$=4~eV
obtained in RPA calculations for metallic iron~\cite {Ferdi}. 

This value for Coulomb parameter (with Hund's exchange parameter
$J$=0.7~eV) was used in Dynamical Mean-Field Theory (DMFT)~\cite {DMFT}
calculations for LaFeAsO \cite {Haule08, DMFT-our, Craco08}. Results of
these works  show iron 3$d$ electrons being in intermediate or strongly
correlated regime, as it is natural to be expected for Coulomb parameter
value $U$=4~eV and Fe-$3d$ band width $\approx$4~eV. 

The most direct way to estimate correlation effects strength in a system
under consideration is to compare the experimental spectra with  densities
of states (DOS) obtained in DFT calculations. For strongly correlated
materials additional features in the experimental photoemission and
absorption spectra appear that are absent in DFT DOS. Those features are interpreted as lower and upper Hubbard
bands. If no Hubbard bands are
observed and DOS obtained in DFT calculations satisfactorily describes the
experimental spectra then the material is in weakly correlated regime. 

LaFeAsO was studied by soft X-ray absorption and emission
spectroscopy~\cite {X-ray}, X-ray absorption (O $K$-edge)
spectroscopy~\cite {exp-U}, and photoemission spectroscopy~\cite
{Malaeb-08}. In all these works the conclusion was that DOS obtained in DFT
calculations gave good agreement with the experimental spectra and the
estimations~\cite {exp-U} for Coulomb parameter value are $U<$1 eV. Such
contradiction with results of DMFT calculations~\cite {Haule08,
DMFT-our, Craco08} using $U$=4~eV shows that \textit {first principles} calculation of 
Coulomb interaction parameter $U$ value for LaFeAsO is needed to determine
the correlation effects strength in this material. Results of such
calculations by constrained DFT calculations are reported in the present
work. We have obtained the value $U<$1~eV for calculation with basis restricted to Fe-$3d$  bands  in agreement with
the estimates from spectroscopy. 

\begin {figure}
\includegraphics [width=0.425\textwidth]{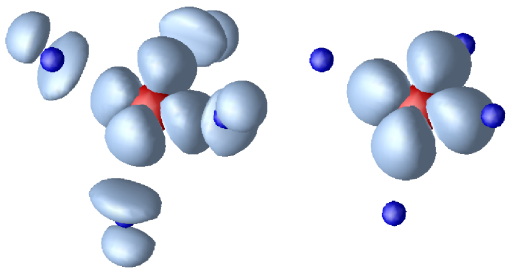}
\caption {(Colour online) Module square of $d_{x^2-y^2}$-like Wannier
function computed for Fe-$3d$ bands only (left panel) and for full set of
O-$2p$, As-$4p$ and Fe-$3d$ bands (right panel). Big sphere in the center
marks Fe ion position and four small spheres around it correspond to As
neighbors.}
\label {fig:WF}
\end {figure}

It is important to note that Coulomb interaction parameter $U$ value
depends on the choice of the model  and, more specifically, on the
choice of the orbital set that is used in the
model. For example, in constrained DFT calculations for high-T$_c$ cuprates
the resulting $U$ value for Cu $d$-shell was found~\cite {cuprate-d}
between 8 and 10~eV. The $U$ value in this range was used in cluster
calculations where all Cu $d$-orbitals and $p$-orbitals of neighboring
oxygens were taken into account and calculated spectra agree well with
experimental data~\cite {cluster-Cu}. However, in one band model, where
only $x^2-y^2$ orbital per copper atom is explicitly included in the
calculations~\cite {cuprate-2D}, the $U$ value giving good agreement with
experimental data falls down to 2.5$\div$3.6~eV, that is 3-4 times smaller
than constrained DFT value. 

The same situation occurs for titanium and vanadium oxides: the $U$ value
from constrained DFT calculations is $\approx$6~eV and cluster calculations
where all $d$-orbitals and $p$-orbitals of neighboring oxygens were taken
into account with $U$ close to this value gave good agreement between
calculated and experimental spectra~\cite {cluster-Ti-V}. However, in the
model where only partially filled $t_{2g}$ orbitals are included, much
smaller $U$ value (corresponding to Slater integral $F^0$=3.5~eV) gives the
results in agreement with experimental data~\cite {V-Ti-dmft}.

\begin {table} [b!]
\caption {\label {tab:UJ} The constrained DFT calculated values of average Coulomb
interaction $\bar{U}$ and Hund's exchange $J$ (eV) parameters for 
$d$-symmetry Wannier functions computed with two different sets of bands
and orbitals.}
\begin {tabular}{l|c|c}
\hline \multicolumn {1}{l}{DFT method} & \multicolumn {1}{|c}{restricted to
Fe-$3d$ bands} & \multicolumn {1}{|c}{full bands set} \\ \hline TB-LMTO-ASA
&  $\bar{U}$=0.49, $J$=0.51 & $\bar{U}$=3.10, $J$=0.81 \\ PWSCF &
$\bar{U}$=0.59, $J$=0.53 &  $\bar{U}$=4.00, $J$=1.02 \\ \hline 
\end {tabular}
\end {table}

It is interesting that such a small $U$ value can be obtained in
constrained DFT calculations for titanates and vanadates where only
$t_{2g}$-orbital occupancies are fixed while all other states
($e_g$-orbitals of vanadium and $p$-orbitals of oxygens) are allowed to relax
in self-consistent iterations~\cite {vanadates-titanates, V-Ti-dmft}. So
the calculation scheme used in constrained DFT (the set of the orbitals
with fixed occupancies) should be consistent with basis set of the model
where the calculated $U$ value will be used.

Another source of uncertainty in constrained DFT calculation scheme is a
definition of atomic orbitals whose occupancies are fixed and energies are
calculated. In some DFT methods, like Linearized Muffin-Tin Orbitals
(LMTO), these orbitals could be identified with LMTO. However, in other DFT
calculation schemes, where plane waves are used as a basis, like in
pseudopotential method one should use more general definition for localized
atomic like orbitals such as Wannier functions~\cite {Wannier37} (WFs). The
practical way to calculate WFs for specific materials using projection of
atomic orbitals on Bloch functions was developed in Ref.~\cite
{MarzariVanderbilt}. 

In Fig.~\ref {fig1} the total and partial DOS for LaFeAsO obtained in  LMTO
calculations are shown. Crystal field splitting for Fe-$3d$ orbitals in
this material is rather weak ($\Delta_{cf}$=0.25~eV) and all five $d$
orbitals of iron form common band in the energy region ($-$2, $+$2)~eV
relative to the Fermi level (see grey region on the bottom panel in
Fig.~\ref {fig1}). There is a strong hybridization of iron $t_{2g}$
orbitals with $p$ orbitals of arsenic atoms which form nearest neighbors
tetrahedron around iron ion. This effect becomes apparent in the energy
interval ($-$3, $-$2)~eV (white region on the bottom panel in Fig.~\ref
{fig1}) where band formed by $p$ orbitals of arsenic is situated. More week
hybridization with oxygen $p$ states reveals in ($-$5.5, $-$3)~eV energy
window (black region on the bottom panel in Fig.~\ref {fig1}).

\begin {figure} 
\includegraphics [width=0.425\textwidth]{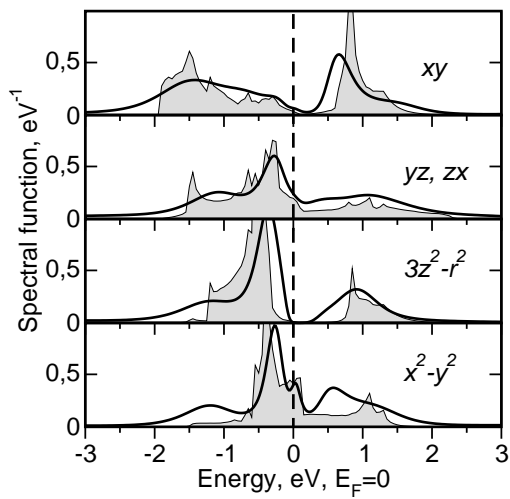} 
\caption {Partial densities of states for  Fe-$3d$ orbitals obtained within
the DFT (filled areas) and LDA+DMFT orbitally resolved spectral functions
for ``restricted basis'' and $F^0$=0.8~eV, $J$=0.5~eV (bold lines).} 
\label {LDA+DMFT-dos}  
\end {figure}

We have calculated Coulomb interaction $U$ and Hund's exchange $J$
parameters for WFs basis set via constrained DFT procedure with fixed
occupancies for WFs of $d$ symmetry. For this purpose we have used two
calculation schemes based on two different DFT methods. One of them
involves linearized muffin-tin orbitals produced by the TB-LMTO-ASA
code~\cite {LMTO}; corresponding WFs calculation procedure is described in
details in Ref.~\cite{WF-LMTO}. The second one is based on the pseudopotential plane-wave method PWSCF, as
implemented in the Quantum ESPRESSO package~\cite {PW} and is described in
Ref.~\cite{WF-PW}. The difference between the results of
these two schemes could give an estimation for the error of $U$ and $J$
determination.

The WFs are defined by the choice of Bloch functions Hilbert space and by a
set of trial localized orbitals that will be projected on these Bloch
functions ~\cite {WF-LMTO}. We performed calculations for two different choices of Bloch
functions and atomic orbitals. One of them includes only bands
predominantly formed by Fe-$3d$ orbitals in the energy window ($-$2,
$+$2)~eV and equal number of Fe-$3d$ orbitals to be projected on the Bloch
functions for these bands. That choice corresponds to the model where only
five $d$-orbital per Fe site are included but all arsenic and oxygen
$p$-orbitals are omitted. Second choice includes all bands in energy window
($-$5.5, $+$2)~eV that are formed by O-$2p$, As-$4p$ and Fe-$3d$ states and
correspondingly full set of O-$2p$, As-$4p$ and Fe-$3d$ atomic orbitals to
be projected on Bloch functions for these bands. That would correspond to
the extended model where in addition to $d$-orbitals all $p$-orbitals are
included too.

\begin {figure}
\includegraphics [width=0.425\textwidth]{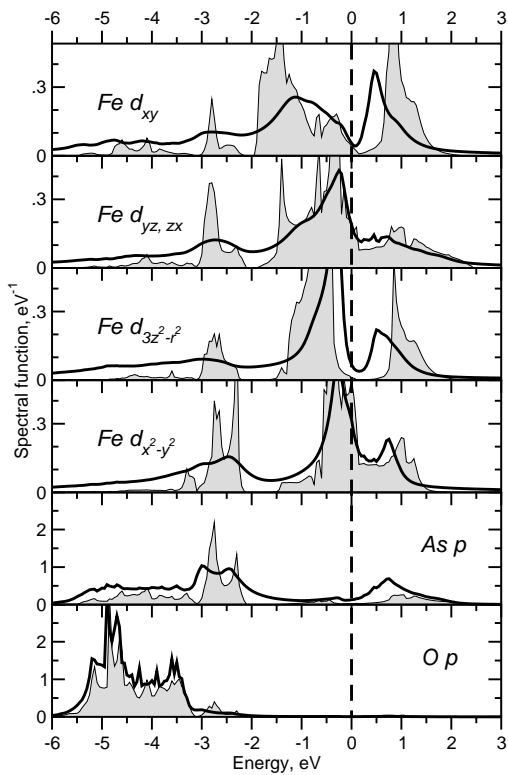}
\caption {Partial densities of states for Fe-$3d$ , As-$4p$ and O-$2p$ states
obtained within the DFT (filled areas) and LDA+DMFT orbitally resolved
spectral functions for ``full basis'' calculations with $F^0$=3.5~eV, $J$=0.81~eV (bold lines).}
\label {LDA+DMFT-dos_p-d} 
\end {figure}

In both cases we obtained Hamiltonian in WF basis that reproduces exactly
bands  formed by Fe-$3d$ states in the energy window ($-$2,
$+$2)~eV (Fig.~\ref {fig1}), but in the second case in addition to that
bands formed by $p$-orbitals in the energy window ($-$5.5, $-$2)~eV will be
reproduced too. However, WFs with $d$-orbital symmetry computed in those
two cases have very different spatial distribution. In Fig.~\ref {fig:WF}
the module square of $d_{x^2-y^2}$-like WF is plotted. While for the case
when full set of bands and atomic orbitals was used (right panel) WF is
nearly pure atomic d-orbital (iron states contribute 99\%), WF computed
using Fe-$3d$ bands only is much more extended in space (left panel). It
has significant weight on neighboring As ions with only 67\% contribution
from central iron atom. 

The physical reason for such effect is $p$-$d$ hybridization that is treated
explicitly in the case where both $p$- and $d$-orbitals are included. In
the case where only Fe-$3d$ bands are included in calculation $p$-$d$
hybridization reveals itself in the shape of WF. Fe-$3d$ bands in the
energy window ($-$2, $+$2)~eV correspond to antibonding combination of
Fe-$3d$ and As-$4p$ states and that is clearly seen on the left panel of
Fig.~\ref {fig:WF}.

\begin {figure}
\includegraphics [width=0.425\textwidth]{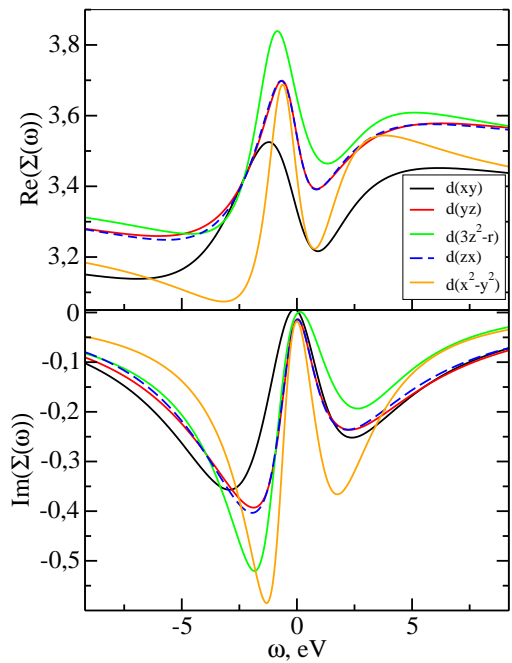}
\caption {(Colour online) Real (upper panel) and imaginary (lower panel)
parts of LDA+DMFT self energy interpolated on real axis with the use of
Pad\'e approximant for ``restricted basis'' and $F^0$=0.8~eV, $J$=0.5~eV.}
\label {d-sigma} 
\end {figure}

The different spatial distribution for two WFs calculated with full and
restricted orbital bases can be expected to lead to different effective
Coulomb interaction for electrons occupying these states. The results of
constrained DFT calculations of the average Coulomb interaction $\bar{U}$
and Hund's exchange $J$ parameters for electrons on WFs computed with two
different set of bands and orbitals (and using two different DFT methods:
LMTO and pseudopotential) are presented in Tab.~\ref {tab:UJ}.

One can see that very different Coulomb interaction strength is obtained
for separate Fe-$3d$ band and full bands set calculations. While the latter
gives value 3$\div$4~eV, close to previously used values~\cite
{Haule08, DMFT-our, Craco08},  Fe-$3d$ band restricted calculation results in
0.5$\div$0.6~eV, that is much smaller but agrees with spectroscopy
estimations~\cite {exp-U}.

The main reason for such a drastic difference between two calculations results  is
very different spatial extension of the two WFs (see Fig.~\ref {fig:WF}):
nearly complete localization on central iron atom for ``full bands set'' WF
(99\%) and only 67\% for ``Fe-$3d$ band set'' WF with 33\% of WF on neighboring As atoms. Another reason
for strong reduction of  calculated $\bar{U}$ value in going from ``full
bands set'' to ``Fe-$3d$ band set'' WF is additional screening via $p$-$d$
hybridization with As-$4p$ band that is situated just below Fe-$3d$ band
(see Fig.~\ref {fig1}). The effect of decreasing of the effective $\bar{U}$
value in several times going from full orbital model to restricted basis
was found previously for high-T$_c$ cuprates ($U$=8$\div$10~eV for full
$p$-$d$-orbitals basis~\cite {cuprate-d} and 2.5$\div$3.6~eV for one-band
model~\cite {cuprate-2D}).

In constrained DFT calculations one obtains an average Coulomb interaction
$\bar{U}$ that can be estimated\cite{anigun} as $\bar{U}=F^0-J/2$ . Hence, Slater
integral $F^0$ can be calculated as $F^0=\bar{U}+J/2$. For
``Fe-$3d$ band set'' WF that gives $F^0$=0.8~eV at $J$=0.5~eV. Coulomb
parameters for ''full basis'' were calculated in the same way using data
from the first row in Tab.~\ref {tab:UJ}, and were taken as $F^0$=3.5~eV
and $J$=0.81~eV. With this set of parameters we performed the
LDA+DMFT~\cite {LDA+DMFT} calculations (for detailed description of the
present computation scheme see Ref.~\cite{WF-LMTO}). The DFT band
structure was calculated within the TB-LMTO-ASA method~\cite {LMTO}.
Crystal structure parameters were taken from Ref.~\cite{Kamihara-08}. 

\begin {figure}
\includegraphics [width=0.405\textwidth]{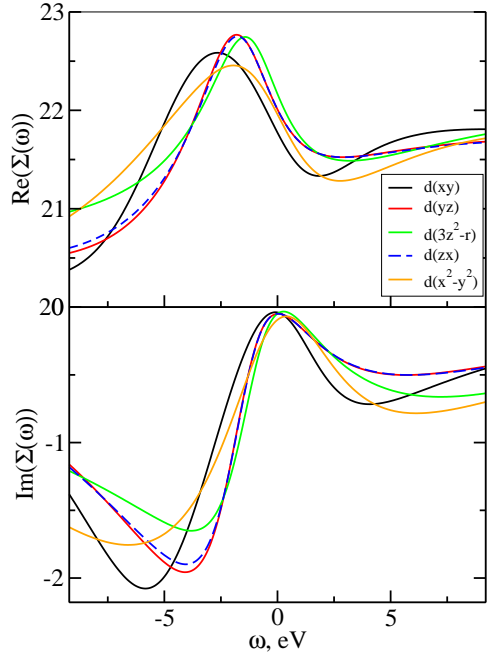}
\caption {(Colour online) Real (upper panel) and imaginary (lower panel)
parts of LDA+DMFT self energy interpolated on real axis with the use of
Pad\'e approximant for ``full basis'' and $F^0$=3.5~eV, $J$=0.81~eV.}
\label {p-d-sigma} 
\end {figure}

The LDA+DMFT calculations were performed for both models: with restricted to Fe-$3d$ states basis
and with full basis including also As-$p$ and O-$p$ states. For the later case
 double counting term $\bar{U}(n_{DMFT}-\frac{1}{2})$ was used to obtain noninteracting Hamiltonian \cite{WF-PW}.
Here $n_{DMFT}$ is  the total number of $d$-electrons obtained
selfconsistently within the LDA+DMFT scheme. The effective impurity model
for the DMFT was solved by the QMC method in Hirsh-Fye algorithm~\cite
{HF86}. Calculations for both cases were performed for the value of inverse
temperature $\beta$=10~$eV^{-1}$. Inverse temperature interval
$0<\tau<\beta$ was divided into 100 slices.  $6 \cdot 10^6$ QMC sweeps were
used in self-con\-sis\-ten\-cy loop within the LDA+DMFT scheme and $12
\cdot 10^6$ of QMC sweeps were used to calculate the spectral functions.

The iron $3d$ orbitally resolved spectral functions obtained within DFT and
LDA+DMFT calculations for ``restricted basis'' with $F^0$=0.8~eV, $J$=0.5~eV are presented in Fig.~\ref
{LDA+DMFT-dos}. The influence of correlation effects on the electronic
structure of LaFeAsO is minimal: there are relatively small changes of peak positions
for $3z^2-r^2$, $xy$ and $x^2-y^2$ orbitals (the shift toward the Fermi
energy) and practically unchanged picture of spectral function distribution
for $yz, zx$ bands. There is no appearance of either Kondo resonance peak
on the Fermi level or Hubbard bands in the energy spectrum with such small
values of $U$ and $J$. 

\begin {figure}
\vspace {5mm}
\includegraphics [width=0.4\textwidth]{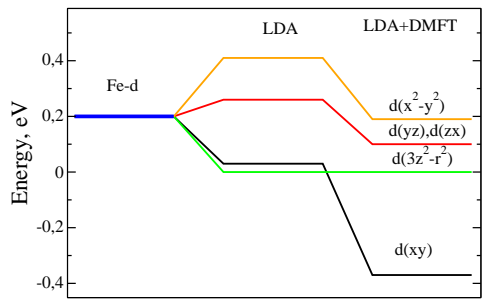}
\caption {(Colour online) Splitting of Fe-$d$ orbitals obtained in LDA and
LDA+DMFT for ``full basis'' and $F^0$=3.5~eV, $J$=0.81~eV.}
\label {split} 
\end {figure}

Results for ``full basis'' LDA+DMFT calculations
are presented in Fig.~\ref{LDA+DMFT-dos_p-d}. Note, that though much larger
Coulomb $U$ and $J$ parameters ($F^0$=3.5~eV, $J$=0.81~eV) were used in this calculation the spectra around the Fermi energy are very similar to those obtained in 
 ``restricted basis'' calculations (Fig.~\ref
{LDA+DMFT-dos}). The general
shape of spectra does not show either Kondo resonance peak on the Fermi
level or Hubbard bands and the features in Fe-$d$ spectral functions below -2 eV correspond to hybridization with  As-$p$ and O-$p$ bands. The reason for such weak correlation effects in spite of relatively large Coulomb repulsion parameters is very strong hybridization of Fe-$d$ orbitals with As-$p$ states (see peaks in Fe-$d$ spectral function in the region -2$\div$-3 eV corresponding to admixture to As-$p$ bands). This hybridization results in additional very effective channel of screening for Coulomb interaction between  Fe-$d$ electrons.

\begin {figure}
\includegraphics [width=0.425\textwidth]{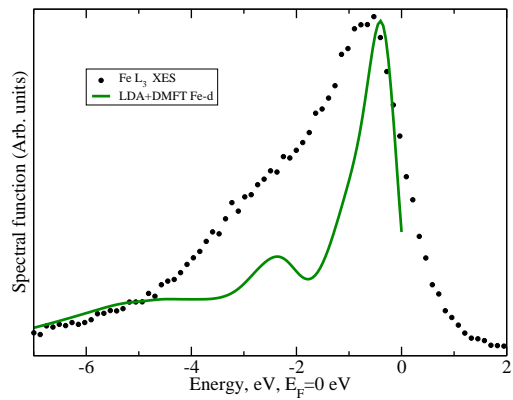}
\caption {(Colour online) Calculated Fe-$d$ LDA+DMFT (``full basis'' and
$F^0$=3.5~eV, $J$=0.81~eV) spectral function (green) and experimental Fe
L$_3$ XES spectrum (black circles) from Ref.~\cite{X-ray}.}
\label {d-spec} 
\end {figure}

This agrees with the results of soft X-ray absorption and emission
spectroscopy study~\cite {X-ray}. It was concluded there that LaFeAsO does
not represent strongly correlated system since Fe $L_3$ X-ray emission
spectra do not show any features that would indicate the presence of the
low Hubbard band or the quasiparticle peak that were predicted by the
LDA+DMFT analysis~\cite {Haule08, DMFT-our, Craco08} with the large
$U$=4~eV. A comparison of the X-ray absorption spectra (O $K$-edge) with
the LDA calculations gave~\cite {exp-U} an upper limit of the on-site
Hubbard $U\approx$1~eV. Photoemission spectroscopy study of LaFeAsO
suggests~\cite {Malaeb-08} that the line shapes of Fe $2p$ core-level
spectra correspond to an itinerant character of Fe $3d$ electrons. It was
demonstrated there that the valence-band spectra are generally consistent
with band-structure calculations except for the shifts of Fe $3d$-derived
peaks toward the Fermi level. Such a shift is indeed observed in our LDA+DMFT results (Fig.~\ref
{LDA+DMFT-dos}).

\begin {figure}
\includegraphics [width=0.425\textwidth]{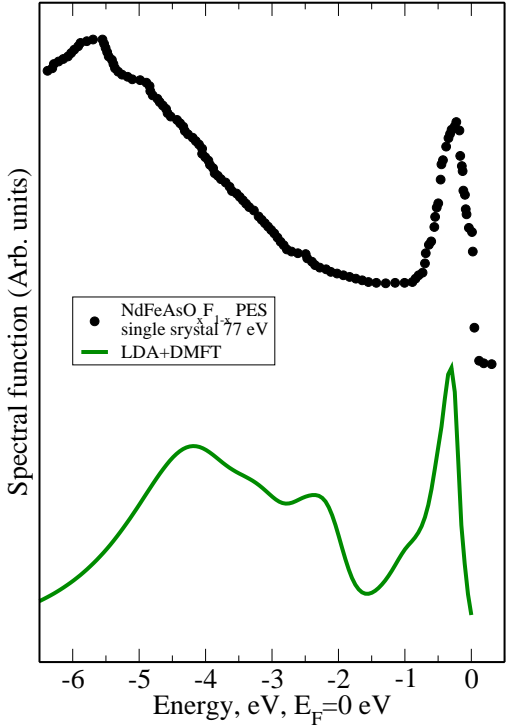}
\caption {(Colour online) Calculated total LDA+DMFT (``full basis'' and
$F^0$=3.5~eV, $J$=0.81~eV) spectral function (green) and experimental
NdFeAsO$_x$F$_{1-x}$ PES spectrum (black circles) from Ref.~\cite{PES}.}
\label {pd-spec} 
\end {figure}

The behavior of real part of self energy near zero frequency
$\Sigma(\omega)|_{\omega \rightarrow 0}$ could give an important
information about band narrowing and renormalization of electron mass in the system under
consideration. Pad\'e approximant~\cite {pade} was used to obtain self
energy on real frequencies. Results for calculation with both bases are presented in
Figs.~\ref {d-sigma}, \ref {p-d-sigma}. The calculated value of
quasiparticle renormalization amplitude
$Z=(1-\frac{\partial\Sigma(\omega)}{\partial\omega}|_{\omega=0)})^{-1}$ was
found to be 0.77, 0.63, 0.71, 0.46 for ``restricted basis'' and 0.56, 0.54,
0.45, 0.56  for ``full basis'' for $d_{xy}$, $d_{yz}$ (or $d_{zx}$),
$d_{3z^2-r^2}$, $d_{x^2-y^2}$ orbitals, respectively. The values for
``restricted basis'' agree well with the effective narrowing of the
LDA+DMFT spectral functions comparing with LDA DOS curves (Fig.~\ref {LDA+DMFT-dos}).
Note, that quasiparticle renormalization amplitude $Z=(1-\frac{\partial\Sigma(\omega)}{\partial\omega}|_{\omega=0)})^{-1}$ from full basis calculations can not be directly used to estimate effective narrowing of the bands. Due to the explicit hybridization with  As-$p$ and O-$p$ states actual narrowing should be weaker.
 Keeping in
mind the above mentioned, we can conclude that both models agree well with
each other. The effective masses $m^*=Z^{-1}$ are 1.3, 1.59, 1.41, 2.17
(``restricted basis'') for $d_{xy}$, $d_{yz}$ (or $d_{zx}$),
$d_{3z^2-r^2}$, $d_{x^2-y^2}$ orbitals, respectively. The $d_{x^2-y^2}$ orbital has the largest effective mass and
exhibits most evident narrowing of LDA spectrum (see Fig.~\ref
{LDA+DMFT-dos}). This orbital has its lobes directed to the empty space
between nearest iron neighbors in  iron plane. Hence it has the weakest
overlapping, the smallest band width, and the largest $U/W$ ratio.

Small effective mass
enhancement value  shows that LaOFeAs belongs to the weakly correlated
systems in contrast to results of the LDA+DMFT calculations~\cite {Haule08}
by Haule {\it et al} where a strongly renormalized low energy band with a
fraction of the original width ($Z\approx$ 0.2-0.3) was found while most of
the spectral weight is transferred into a broad Hubbard band at the binding
energy $\approx$4 eV. Authors of Ref.~\cite{Haule08} report that
"slightly enhanced Coulomb repulsion ($U$= 4.5 eV) opens the gap" so that
the system is in strongly correlated regime on the edge of metal-insulator
transition.

In LaFeAsO the iron ion is in tetrahedral coordination of four As ions with
slight tetragonal distortion of the tetrahedron. In tetrahedral symmetry
group T$_d$ five $d$-orbitals should be split by crystal field~\footnote{
The coordinate axes $x,y$ in LaOFeAs crystal structure are rotated on 45
degrees from the standard tetrahedral notation so that $xy$ and $x^2-y^2$
orbitals are interchanged.}
on low-energy doubly degenerate set $3z^2-r^2$, $xy$  corresponding to
irreducible representation $e_g$ and high-energy triply degenerate set
$x^2-y^2$, $xz$, $yz$ for representation $t_{2g}$. We have calculated
Wannier functions energy and have found that $t_{2g}$--$e_g$ crystal field
splitting parameter is very small $\Delta_{cf}\approx$0.25~eV. The slight
tetragonal distortion of the tetrahedron gives additional  splitting of
$t_{2g}$ and $e_g$ levels with the following values for orbital energies
(energy of the lowest $3z^2-r^2$ orbital is taken as a zero):
$\varepsilon_{3z^2-r^2}$=0.00~eV, $\varepsilon_{xy}$=0.03~eV,
$\varepsilon_{xz,yz}$=0.26~eV, $\varepsilon_{x^2-y^2}$=0.41~eV. Correlation
effects lead not only to narrowing of the bands but also to substantial shift of  Fe-$d$ orbitals energies. For
``full basis'' calculation adding of $Re(\Sigma(0))$ to the LDA orbital
energies results in $\varepsilon_{3z^2-r^2}$=0.00~eV,
$\varepsilon_{xy}$=-0.37~eV, $\varepsilon_{xz,yz}$=0.10~eV,
$\varepsilon_{x^2-y^2}$=0.20~eV (see Fig.~\ref {split}). Note that the
actual splitting should be smaller due to $p$-$d$ hybridization.

Comparisons of spectral function calculated within LDA+DMFT for full basis
set with various experimental spectra are presented in Figs.~\ref {d-spec},
\ref {pd-spec}. One can see a good agreement in comparison of calculated Fe-$d$ spectral function   with Fe
L$_3$ XES spectrum (black circles) from Ref.~\cite{X-ray}. The shoulder in experimental curve
near -2.5 eV correspond to the low energy peak in calculated spectrum that appears due to
strong hybridization between Fe-$d$ and As-$p$ states (see also Fig.~\ref
{fig1}). Note that this features in calculated curve is absent in the calculations
with restricted basis (see orbital resolved spectra in Fig.~\ref
{LDA+DMFT-dos}).  

The total spectrum (Fig.~\ref {pd-spec}) is in reasonable agreement with
experimental photoemission data. The sharp peak at the Fermi energy corresponds to partially filled Fe-$d$ band. Broad bands between -2 and 6 eV represent oxygen and arsenicum $p$ bands.

In conclusion, we have calculated the values of Coulomb interaction parameters $U$ and $J$ via constrained
DFT procedure in the basis of Wannier functions. For minimal model including only Fe-$3d$
orbitals and bands we have obtained Coulomb parameters $F^0$=0.8~eV, $J$=0.5~eV. For
``full basis'' calculations  Coulomb parameters are $F^0$=3.5~eV, $J$=0.8~eV. 
The LDA+DMFT calculation for both models with calculated  parameters
results in weakly correlated nature of iron $d$ bands in this compound.
This conclusion is supported by  spectroscopic investigations
of this material. As the choice of the orbitals to construct the effective Hubbard model from
LDA results is to some extent arbitrary, showing that different choices
leading to largely different interaction parameters yield similar
physical observables provides an important consistency test of the whole
procedure.

Support by the Russian Foundation for Basic Research under Grant No.
RFFI-07-02-00041, Civil Research and Development Foundation together with
Russian Ministry of science and education through program Y4-P-05-15,
Russian president grant for young scientists MK-1184.2007.2,   President of Russian Federation fund of support for scientific schools grant 1941.2008.2 and Dynasty
Foundation is gratefully acknowledged. J.K. acknowledges the support of SFB
484 of the Deutsche Forschungsgemeinschaft.

\end {document}